\documentclass[twocolumn,prl,aps,preprintnumbers,
superscriptaddress,amsmath,amssymb,amsfonts,nofootinbib,showpacs,floatfix]{revtex4} 

\usepackage{graphicx}

\newcommand{\g}{\gamma}
\newcommand{\nn}{\nonumber}
\newcommand{\azmom}{M_{N}^{\gamma j}}
\newcommand{\de}{d}
\newcommand{\bm}{\boldsymbol}

\allowdisplaybreaks[2]

\begin{document}
\title{
The Sivers single-spin asymmetry in photon-jet production
}

\author{Alessandro Bacchetta}
\affiliation{
Theory Group, Deutsches Elektronen-Synchroton DESY,
22603 Hamburg, Germany
}

\author{Cedran Bomhof}
\affiliation{
Dept.\ of Physics and Astronomy, Vrije Universiteit Amsterdam, 1081 HV Amsterdam, The Netherlands
}

\author{Umberto D'Alesio}
\affiliation{
INFN, Sezione di Cagliari and Dipartimento di Fisica, 
Universit{\`a} di Cagliari,
09042 Monserrato, Italy
}

\author{Piet J. Mulders}
\affiliation{
Dept.\ of Physics and Astronomy, Vrije Universiteit Amsterdam,
1081 HV Amsterdam, The Netherlands
}

\author{Francesco Murgia}
\affiliation{
INFN, Sezione di Cagliari and Dipartimento di Fisica, 
Universit{\`a} di Cagliari,
09042 Monserrato, Italy
}

\begin{abstract}
We study a weighted asymmetry in the azimuthal distribution of
photon-jet pairs produced in the process $p^\uparrow p \to \gamma$ jet
$X$ with a transversely polarized proton. We focus on the contribution
of the Sivers effect only, considering experimental
configurations accessible at RHIC-BNL. We show that
predictions for the asymmetry, obtained in terms of gluonic-pole cross
sections calculable in perturbative QCD, can be tested and clearly
discriminated from those based on a generalized parton model,
involving standard partonic cross sections. Experimental measurements
of the asymmetry will therefore test our present understanding
of single-spin asymmetries.
\end{abstract}

\preprint{DESY 07-028}


\maketitle


Single-spin asymmetries (SSA), particularly in processes 
with transversely polarized targets, 
have been measured in proton-proton collisions
$p^{\uparrow} p \to \pi X$ (see, e.g., \cite{Adams:1991cs})
and
semi-inclusive deep inelastic scattering 
(SIDIS), $\ell p^{\uparrow} \to \ell' \pi
X$~\cite{Airapetian:2004tw}.
Different theoretical approaches have been adopted to interpret these
asymmetries and make predictions for other processes. 
In this paper, we make a clear-cut prediction for a simple
process using the color-gauge-invariant QCD formalism (see, e.g.,
\cite{Qiu:1998ia,Bacchetta:2005rm,Kouvaris:2006zy}) and compare it with the
frequently used generalized parton model (see, e.g., \cite{Anselmino:2005sh}).

In general, nonvanishing SSA require 
the interference between scattering amplitudes with different
phases.
Possible sources of phase shifts
are
initial- or final-state color interactions~\cite{Brodsky:2002cx}. When
describing high-energy processes, these color interactions can be included
in parton distribution functions (PDFs). 
In standard gauges, they can be 
identified with the Wilson lines  
required to make the PDFs
 gauge invariant (see, e.g., \cite{Ji:2002aa}).

The form of the Wilson line is fixed by the hard
part of the scattering process and thus process-dependent.
For instance, 
in SIDIS the Wilson line is 
future-pointing (it arises from gluon interactions with the outgoing
quark), 
while in the Drell-Yan (DY) process ($p p^{\uparrow} \to l \bar{l} X$) 
the Wilson line is past-pointing (it arises
from gluon interactions with the incoming antiquark)~\cite{Collins:2002kn}.
This
has a striking consequence for single-spin asymmetries.
In the color-gauge-invariant approach, 
the asymmetries in DY have exactly the {\em opposite} sign 
compared to the generalized parton model expectation.
This sign difference is a fundamental QCD prediction
and its experimental verification would be crucial 
to confirm the validity of our present conceptual framework for analyzing hard
hadronic reactions~\cite{Efremov:2004tp,Bomhof:2007su}. 

When considering a process different from SIDIS and DY, for instance 
$p^{\uparrow} p \to$ hadrons,
the Wilson line structure becomes more 
intricate~\cite{Bomhof:2004aw}. First of all, several partonic QCD processes
contribute; secondly, each process has colored partons both in the initial and
the final state, resulting in a competing effect of future- and past-pointing
Wilson lines. It is therefore more challenging to derive clear-cut predictions
for the sign of the SSA in these processes~\cite{Bomhof:2007su}.

In this letter we shall consider hadronic 
production of a photon and a jet in opposite hemispheres.
This is the simplest case to test the formalism in processes with 
QCD hard scattering. 
After describing the kinematics of the process, we define 
a suitable weighted azimuthal asymmetry that contains 
the Sivers function~\cite{Sivers:1990cc}.
We then present 
quantitative studies in a specific kinematical region and predict the sign of
the asymmetry, which turns out to be opposite to the generalized parton 
model expectation,
 based
on SIDIS results.
The experimental confirmation of this
prediction 
has the same
significance as measuring the relative sign difference of the Sivers effect in 
SIDIS and the Drell-Yan process, and has 
the advantage that the cross-section for
photon production is larger than for Drell-Yan.

\begin{figure}
\includegraphics[width=5cm]{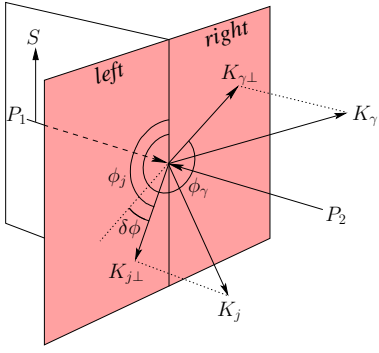}
\caption{Azimuthal angles involved in the process. The vectors $K_{\g \perp}$,
  $K_{j \perp}$ lie on the plane perpendicular to $P_1$.}
\label{f:angles}
\end{figure}

The process under consideration is (see also \cite{Vogelsang:2005cs})
\begin{equation} 
p^{\uparrow}(P_1)+ p(P_2)\to \gamma(K_{\g})+{\rm jet}(K_j)+X.
\end{equation} 
This process 
is
similar to $p^{\uparrow} p\to {\rm jet}\,{\rm jet}\,X$ studied in
\cite{Boer:2003tx,Bacchetta:2005rm,Bomhof:2007su}, 
to polarized DY (see, e.g., \cite{Boer:1999mm}),
and to
$p^{\uparrow} p\to \gamma\,X$ (see, e.g.,
\cite{Schmidt:2005gv}).  
We fix the $z$ direction along $\bm{P}_1$ in the
center-of-mass frame (c.m.). We use the
pseudorapidities  $\eta_i = -\ln \tan(\theta_i/2)$, where $\theta_i$
is the c.m.\ polar angle of the outgoing photon or jet. 
The components of
the outgoing momenta perpendicular to $\bm{P}_1$ are 
denoted as $\bm{K}_{i \perp}$. We introduce the variables $x_{i \perp} = 2
|\bm{K}_{i \perp}|/\sqrt{s}$ 
and the azimuthal angles 
(see Fig.~\ref{f:angles})~\cite{Bacchetta:2004jz} 
\begin{equation} 
\begin{split}
\cos \phi_i &= 
  \frac{(\hat{\bm P}_1\times{\bm S})}{|\hat{\bm P}_1\times{\bm S}|}
  \cdot \frac{(\hat{\bm P}_1\times{\bm K}_i)}{|\hat{\bm P}_1
     \times{\bm K}_i|},\\
\sin \phi_i &= 
  \frac{(\hat{\bm P}_1 \times {\bm S}) \cdot {\bm K}_i}
        {|\hat{\bm P}_1 \times{\bm S}|\,|\hat{\bm P}_1\times{\bm
            K}_i|}, 
\end{split}
\end{equation} 
with $\hat{\bm P}_1={\bm P}_1/|{\bm P}_1|$, where all vectors refer to the c.m.\
(or to any frame connected to the c.m.\ 
by a boost along $\hat{\bm P}_1$). 
Finally, we introduce the vector  
$
r_{\perp} = K_{\g \perp} +  K_{j \perp},
$
and the angle 
$\delta\phi=\phi_j-\phi_\gamma-\pi$.
We focus our attention on 
the case in which $|\bm{r}_{\perp}| \ll |\bm{K}_{\g \perp} -  \bm{K}_{j \perp}|$,
i.e., when the photon and the jet are approximately back-to-back in the
transverse plane. We retain only leading-order contributions 
in an expansion in $|\bm{r}_{\perp}|/|\bm{K}_{\g \perp}|$. In particular, this implies
that 
$x_{\g \perp}=x_{j \perp} \equiv x_\perp$. For comparison's sake, 
we will consistently make the same approximation in the
generalized parton
model~\cite{Anselmino:2005sh}.

We now consider the following azimuthal moment~\cite{Bacchetta:2005rm}
\begin{align}
\label{e:azmoment}
&\azmom(\eta_\gamma,\eta_j,x_\perp) = 
\\&\frac{\int\! \de \phi_j\, \de \phi_\g 
\frac{2 |\bm{K}_{\g \perp}|}{M} \sin(\delta\phi)\cos(\phi_\gamma)\,\frac{\de \sigma}{\de \phi_j\, \de \phi_\g}}
{\int\! \de \phi_j\, \de \phi_\g 
\,\frac{\de \sigma}{\de \phi_j\, \de \phi_\g}} \equiv -\frac{A+B}{C}.
\nonumber
\end{align}
We expect the above integral to be dominated by the
small-$\delta\phi$ region. 
Note that a positive value for this moment means that the sum of the photon
and jet transverse momenta, $\bm{r}_{\perp}$, has a preference to lie on the right
side of the transverse plane (as defined in Fig.~\ref{f:angles}), i.e.,
the photon--jet pair has
a preference to go to the right.

\begin{figure}
\begin{tabular}{ccc}
\includegraphics[width=2.4cm]{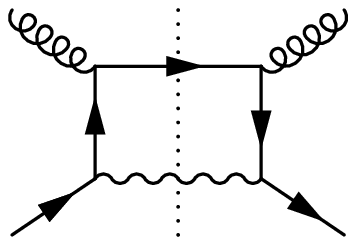}
&\phantom{aaaa}&
\includegraphics[width=2.4cm]{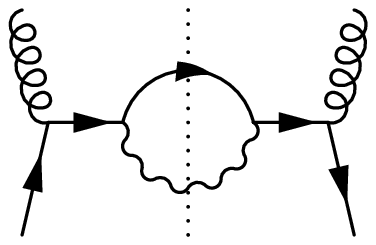}
\\
\\
\includegraphics[width=2.4cm]{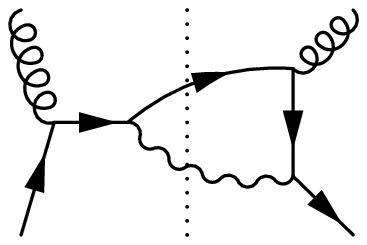}
&&
\includegraphics[width=2.4cm]{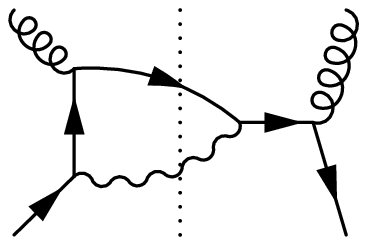}
\end{tabular}
\caption{Cut diagrams for $qg{\to}\gamma q$ scattering.
}
\label{f:qg2qp}
\end{figure}
\begin{figure}
\begin{tabular}{ccc}
\includegraphics[width=2.4cm]{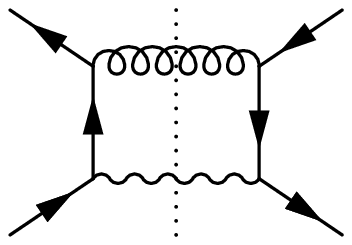}
&\phantom{aaaa}&
\includegraphics[width=2.4cm]{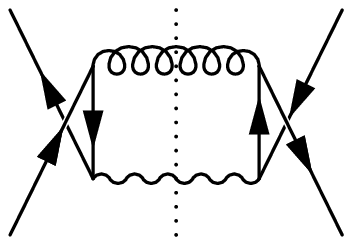}
\\
\\
\includegraphics[width=2.4cm]{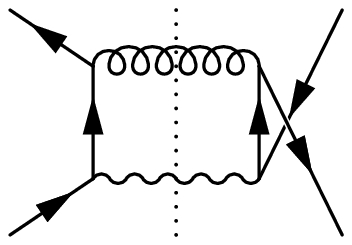}
&&
\includegraphics[width=2.4cm]{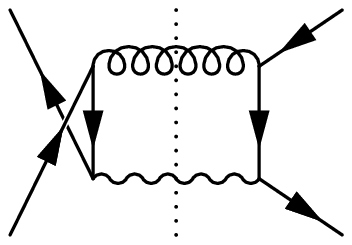}
\end{tabular}
\caption{
Cut diagrams for $q\bar q{\to}\gamma g$ scattering.
}
\label{f:qq_2pg}
\end{figure}
In terms of PDFs and partonic hard cross sections,
the denominator of the above moment can be interpreted as
\begin{equation}
\begin{split}
C &= x_{\perp} x_1 x_2 \,  \textstyle{\sum_{q}}\Bigl[ 
        f_{1}^{g}(x_1)\,f_1^{q}(x_2)\, \de \hat{\sigma}_{g q{\scriptscriptstyle \to} \g q}
        + f_{1}^{q}(x_1)
\\ 
 &\quad \times
        \Bigl(f_1^{\bar{q}}(x_2)\, \de \hat{\sigma}_{q \bar{q} {\scriptscriptstyle \to} \g g}
        +f_1^{g}(x_2)\, \de \hat{\sigma}_{q g {\scriptscriptstyle \to} \g q} \Bigr)\Bigr],
\label{e:denom}
\end{split}
\end{equation}  
where $f_1$ are the unpolarized PDFs and 
the sum runs over quarks and antiquarks.
The standard partonic cross sections appearing in Eq.~(\ref{e:denom}) can be
obtained from the cut diagrams of 
Figs.~\ref{f:qg2qp} and \ref{f:qq_2pg} and
read
\begin{align}
\de \hat{\sigma}_{q \bar{q} {\scriptscriptstyle \to} \g g} &=
\frac{\pi \alpha \alpha_S
  e_q^2}{\hat{s}^2}\,\frac{N_c^2-1}{N_c^2}\,
        \biggl(\frac{\hat{u}}{\hat{t}}+\frac{\hat{t}}{\hat{u}} \biggr), 
\\
\de \hat{\sigma}_{q g {\scriptscriptstyle \to} \g q} &=
\frac{\pi \alpha \alpha_S
  e_q^2}{\hat{s}^2}\,\frac{1}{N_c}\,
        \biggl(-\frac{\hat{t}}{\hat{s}}-\frac{\hat{s}}{\hat{t}} \biggr),
\\
\de \hat{\sigma}_{g q{\scriptscriptstyle \to} \g q} &=
\frac{\pi \alpha \alpha_S
  e_q^2}{\hat{s}^2}\,\frac{1}{N_c}\,
        \biggl(-\frac{\hat{u}}{\hat{s}}-\frac{\hat{s}}{\hat{u}} \biggr),
\\
\de \delta\hat{\sigma}_{q^{\uparrow} \bar{q}^{\uparrow} {\scriptscriptstyle \to} \g g} &=
\frac{\pi \alpha \alpha_S
  e_q^2}{\hat{s}^2}\,\frac{N_c^2-1}{N_c^2}\,(-2),
\end{align} 
where the last term has been included for later use.
The momentum fractions $x_1$ and $x_2$ and 
the partonic Mandelstam variables can be expressed as
\begin{gather}
\begin{align} 
x_1&= \frac{x_\perp}{2} (e^{\eta_\g} + e^{\eta_j}),
&x_2&= \frac{x_\perp}{2} (e^{-\eta_\g} + e^{-\eta_j}), 
\end{align}
\\
\begin{align}
\hat{s}&= x_1\, x_2\, s,
&
-\frac{\hat{t}}{\hat{s}} &\equiv y = \frac{1}{e^{\eta_\gamma - \eta_j}{+}1},
&
-\frac{\hat{u}}{\hat{s}} &=1-y. 
\end{align} 
\end{gather}

The contributions $A$ and $B$  in Eq.~\eqref{e:azmoment} are given by
\begin{align} 
A &= x_{\perp} x_1 x_2 \,\textstyle{\sum_{q}}\Bigl[f_{1T}^{\perp (1)g_d}(x_1)\,f_1^{q}(x_2)\, 
                \de \hat{\sigma}_{[g] q {\scriptscriptstyle \to} \g q}^{(d)} 
\nn \\
&\quad + f_{1T}^{\perp (1)g_f}(x_1)\,f_1^{q}(x_2)\,
                \de \hat{\sigma}_{[g] q{\scriptscriptstyle \to} \g q}^{(f)}
\label{e:num}
 \\ 
 &\quad 
+ f_{1T}^{\perp (1) q}(x_1)
        \Bigl(f_1^{\bar{q}}(x_2) \de \hat{\sigma}_{[q] \bar{q} {\scriptscriptstyle \to} \g g}
        +f_1^{g}(x_2) \de \hat{\sigma}_{[q] g {\scriptscriptstyle \to} \g q} \Bigr)\Bigr],
\nn
\\
\begin{split}
B &= x_{\perp} x_1 x_2 \, \textstyle{\sum_{q}} h_{1}^{q}(x_1) h_{1}^{\perp (1)\bar{q}}(x_2)\, 
        \de \delta \hat{\sigma}_{q^{\uparrow} [\bar{q}]^{\uparrow} {\scriptscriptstyle \to} \g g},
\label{e:numB}
\end{split}
\end{align} 
where the transversity function ($h_1$), and
the first transverse moments of the Sivers function 
($f_{1T}^{\perp (1)}$) and of the Boer-Mulders function ($h_{1}^{\perp
  (1)}$)~\cite{Boer:1997nt} appear.
Note that there are two different gluon Sivers functions, corresponding to 
two distinct ways to construct color-singlet
three-gluon matrix elements, using the symmetric $d^{abc}$ and
antisymmetric $f^{abc}$ structure constants of $SU(3)$,
respectively~\cite{Bomhof:2006ra}.
The modified partonic cross sections  in the above equations are
the so-called gluonic-pole cross
sections~\cite{Bacchetta:2005rm}. They are 
gauge-invariant sums of Feynman
diagrams weighted with  multiplicative prefactors, 
called gluonic-pole strengths. These can be
computed using the procedure outlined in
\cite{Bacchetta:2005rm,Bomhof:2006ra} and 
are a direct consequence of the presence of the Wilson lines. They
generalize  
the $\pm 1$ prefactors appearing in SIDIS and DY and are entirely
determined by the color topology of the involved QCD partonic diagram.
Gluonic-pole cross
sections are particularly simple in the 
case considered here because
the photon is colorless and all the subprocesses in Fig.~\ref{f:qg2qp} have
the same color structure, and so do all the subprocesses in
Fig.~\ref{f:qq_2pg}. Therefore, the inclusion of the Wilson lines results
simply in common prefactors: 
\begin{align}
\de \hat\sigma_{[q]\bar q{\scriptscriptstyle \to} \g g}
   & =\frac{N_c^2{+}1}{N_c^2{-}1}\,\de \hat\sigma_{q\bar q{\scriptscriptstyle \to} \g g},
\label{e:plus}
\\
\de \hat\sigma_{[q]g{\scriptscriptstyle \to} \g q}
   & =-\frac{N_c^2{+}1}{N_c^2{-}1}\, \de \hat\sigma_{qg{\scriptscriptstyle \to} \g q},
\label{e:minus}
\\
\de \hat\sigma_{[g] q{\scriptscriptstyle \to} \g q}^{(d)}
   & = \de \hat\sigma_{gq{\scriptscriptstyle \to} \g q},
\qquad
\de \hat\sigma_{[g]q{\scriptscriptstyle \to} \g q}^{(f)}
    = 0,
\\
\de \delta \hat\sigma_{q^{\uparrow} [\bar{q}]^{\uparrow} {\scriptscriptstyle \to} \g g}
   & =\frac{N_c^2{+}1}{N_c^2{-}1}\, \de \delta \hat\sigma_{q^{\uparrow} \bar{q}^{\uparrow} {\scriptscriptstyle \to} \g g}.
\end{align} 
The most significant difference between the standard partonic cross sections
and the gluonic-pole cross sections is the minus sign in
Eq.~(\ref{e:minus}).
This sign, entirely due to the color structure of the partonic process, is
a straightforward consequence of QCD.
In particular, the different
signs in Eq.~(\ref{e:plus}) and Eq.~(\ref{e:minus}) are due
to the fact that we have in the first case an incoming (anti)quark and an
outgoing gluon and in the
second case an incoming gluon and an outgoing (anti)quark. In the large-$N_c$
limit, in the first case the color flows from the incoming
quark into the final state as in SIDIS,
while in the second case the color flows back into the initial state as in DY.

\begin{figure}
\includegraphics[width=6.5cm]{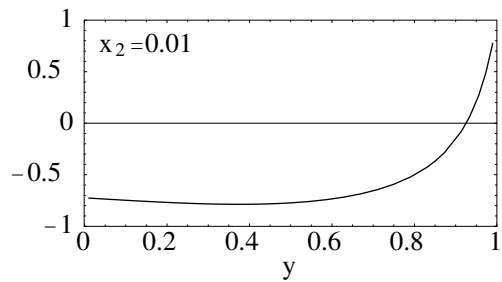}
\caption{Behavior of the last term of Eq.~(\ref{e:azmomapprox}) at $x_2=0.01$.}
\label{f:shape}
\end{figure}

To have an idea of the impact of the negative sign in Eq.~\eqref{e:minus}, 
before presenting a detailed numerical study of Eq.~\eqref{e:azmoment}, we
discuss 
a simplified situation. 
We consider
the high-$x_1$
region, 
where the  sea-quark contributions in the polarized proton can be
neglected.  
We also neglect the Boer-Mulders function
and assume
a symmetric-sea scenario, i.e., $f_1^{\bar{d}}\approx f_1^{\bar{u}}\equiv
f_1^{\bar{q}}$.  
In this way
the azimuthal moment we are studying can be written as
\begin{align}
\label{e:azmomapprox}
&\azmom
\approx -
\frac{5}{4}\,
\frac{4\,f_{1T}^{\perp (1) u}(x_1)+f_{1T}^{\perp (1) d}(x_1)}{4\,f_1^{u}(x_1)
+\,f_{1}^{d}(x_1)}\,
\\
&\times
\frac{f_1^{\bar{q}}(x_2)\,\bigl(1-2y+2 y^2\bigr) -\frac{3}{8}\,f_1^{g}(x_2)\,(1-y)\,\bigl(1+y^2\bigr)}
{f_1^{\bar{q}}(x_2)\,\bigl(1-2y+2 y^2\bigr) +
  \frac{3}{8}\,f_1^{g}(x_2)\,(1-y)\,\bigl(1+y^2\bigr)}.
\nonumber
\end{align}
We first analyze the behavior of the last term of the azimuthal
moment as a function of the two variables $x_2$ and $y$. We use the GRV98LO
set of PDFs~\cite{Gluck:1998xa} 
at the indicative scale 
$\hat{s}=200$ GeV$^2$. The result is
plotted in Fig.~\ref{f:shape} for $x_2=0.01$. The behavior is similar for any
other value of $x_2$.
In most of the $x_2$ and $y$ space this coefficient is large and 
{\em negative},
due to the dominance of the gluon distribution function over the sea quark
one. 
The result holds true for any set of PDFs at any reasonable scale.
We emphasize that 
if standard partonic cross sections were used, this coefficient
should be equal to one.  
Parameterizations of the Sivers
distribution functions indicate that $f_{1T}^{\perp (1) u}$ is negative and
$f_{1T}^{\perp (1) d} \approx - f_{1T}^{\perp (1) u}$~\cite{Vogelsang:2005cs,Anselmino:2005ea,Collins:2005ie}. Therefore, we expect
the azimuthal moment to be {\em negative}, i.e., we expect the photon-jet pair
to go preferably to the left,
 {\em opposite} 
to the expectation of the generalized parton model, which uses 
standard partonic cross sections both in Eq.~\eqref{e:denom} and 
Eqs.~\eqref{e:num}, 
\eqref{e:numB}. 

To confirm the above expectation, we perform a more detailed numerical study
of Eq.~\eqref{e:azmoment}. We use the unpolarized PDFs at
the scale 
$\hat{s}= x_1 x_2 s$. For the up and down Sivers function we use the results
of the fit of \cite{Anselmino:2005ea}. 
We saturate the transversity distribution
function using the Soffer bound~\cite{Soffer:1995ww} with the
GRSV2000~\cite{Gluck:2000dy} polarized PDFs. 
For the gluon Sivers function and
the Boer-Mulders function we saturate the positivity 
bound~\cite{Bacchetta:1999kz}
\begin{equation*} 
\bigl|f_{1T}^{\perp (1)g}(x)\bigr|\le \int d^2 \bm{p}_T\,
\frac{|\bm{p}_T|}{2 M}\,f_1^{g}(x,\bm{p}_T^2)\approx 
\frac{\langle |\bm{p}_T| \rangle}{2 M}\, f_1^{g}(x), 
\end{equation*}
which holds also for $h_{1}^{\perp (1)\bar{q}}$. We use 
$\langle |\bm{p}_T| \rangle=0.44$ GeV~\cite{Anselmino:2005nn}.
We neglect
the sea-quark Sivers functions.

\begin{figure}
\includegraphics[width=6.3cm]{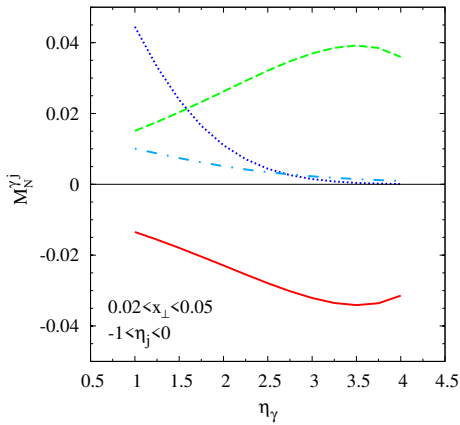}
\caption{Prediction for the azimuthal moment $\azmom$
  at $\sqrt{s} = 200$ GeV, as a
  function of $\eta_\gamma$,
  integrated over $-1 \le \eta_j \le 0$ and $0.02 \le x_{\perp} \le 0.05$. 
  Solid
  line: using gluonic-pole cross sections. Dashed line: using standard
  partonic cross sections. Dotted line: maximum 
  contribution from the gluon Sivers function (absolute value). 
  Dot-dashed line: maximum
  contribution from the Boer-Mulders function (absolute value).}
\label{f:azmoment}
\end{figure}

In order to emphasize the effect of the sign change in Eq.~\eqref{e:minus}, we
need to select small values of $y$, where the partonic subprocesses 
$q\bar q\to \g g$ and $q g\to \g q$ dominate. Moreover, in
order to have a sizeable quark Sivers function, we need to select 
$x_1\sim 0.2-0.3$. These two conditions can be fulfilled by choosing large 
positive values for $\eta_\gamma$ and small or negative values for $\eta_j$. In
Fig.~\ref{f:azmoment}  we present our estimate for $\azmom$ at $\sqrt{s} =
200$ GeV (RHIC kinematics), as a
  function of $\eta_\gamma$,
  integrated over $-1 \le \eta_j \le 0$ and $0.02 \le x_{\perp} \le 0.05$. 
The solid line
represents our prediction when taking into account only the up and down
quark Sivers
function. 
The maximum contributions from the gluon Sivers function and 
the Boer-Mulders function (dotted and dash-dotted lines) turn out to be
negligible at high 
$\eta_\gamma$. 
Thus, we can robustly 
predict $\azmom$ to be negative in this
kinematical regime. 
In contrast, the generalized parton model (dashed line in 
Fig.~\ref{f:azmoment}) predicts the opposite sign.

In conclusion, we have examined the azimuthal moment $\azmom$, defined in
Eq.~\eqref{e:azmoment}, for the process $p^{\uparrow} p\to \gamma\, {\rm jet}\,
X$. We have shown that in the kinematical regime of large and positive photon
pseudorapidities and negative jet pseudorapidities, the moment is dominated by
the quark Sivers function 
combined with the gluon unpolarized distribution function. 
The involved partonic subprocess is $q g \to \g q$. The two
functions have to be convoluted with a gluonic-pole cross section instead of a
standard partonic cross section, to take into account the presence of
past-pointing 
 and future-pointing
 Wilson
lines arising from gluon interactions with the incoming gluon 
and the outgoing
quark, respectively. The color structure of QCD implies that
the gluonic-pole cross section for $q g \to \g q$ 
is equal to $-5/4$ times the standard
partonic cross section. This leads to the robust 
prediction of a negative sign for
the azimuthal moment $\azmom$ in the considered kinematical regime, opposite
to the expectation of the generalized parton model, obtained using 
standard partonic cross sections. The
experimental measurement of $\azmom$, 
possible at RHIC, will therefore be of
crucial importance to deepen our present understanding of
single-spin asymmetries. 

This work is part of the EU Integrated
Infrastructure Initiative Hadron Physics (RII3-CT-2004-506078). The
work of C.~B.\ is supported by the Dutch Foundation for Fundamental Research of
Matter (FOM) and the Dutch National Organization for Scientific Research (NWO).

\vspace{-2mm}

\bibliographystyle{myrevtex2}
\bibliography{mybiblio}

\end{document}